# Roman CCS White Paper

**Title: Isolated Stellar-Mass Black Holes*:* Strategy to Improve the Efficiency and Robustness of Detection with *Roman***

**Roman Core Community Survey:** *Galactic Bulge Time Domain Survey*
**Scientific Categories:** *stellar physics and stellar types; stellar populations and the interstellar medium*
**Additional scientific keywords:** *Black Holes, Massive Stars*

**Submitting Author:**
Name: Kailash Sahu
Affiliation: STScI
Email: ksahu@stsci.edu

**List of contributing authors** (including affiliation and email):
Kailash C. Sahu, Space Telescope Science Institute, 3700 San Martin Drive, Baltimore, MD 21218, USA. Email: ksahu@stsci.edu

Sedighe Sajadian, Department of Physics, Isfahan University of Technology, Isfahan 84156-83111, Iran. Email: s.sajadian@iut.ac.ir

**Abstract:**

*Roman* telescope provides the best opportunity to detect a large number of Isolated Stellar-Mass Black Holes (ISMBHs) through microlensing. *Roman* will not only detect long-duration microlensing events caused by ISMBHs, but will also measure the deflections caused by the ISMBHs, which can be used to estimate their masses. Recently, Sajadian and Sahu (2023) studied the efficiency of detecting ISMBHs by *Roman* through simulation of a large ensemble of such events. They estimated the resulting errors in the physical parameters of the lens objects, including their masses, distances, and proper motions through calculating Fisher and Covariance matrices. Their simulation shows that the ~2.3-year time gap between *Roman*'s first three and the last three seasons not only lowers the efficiency of detection, but also makes the solutions degenerate. *We recommend a small amount of additional observations –one hour of observations every 10 to 20 days when the Bulge is observable during the large time gap– which is equivalent to a total of about one to two additional days of observations with Roman.* This small amount of additional observations will greatly improve the efficiency and robustness of detection of ISMBHs, and provide firm estimates of their masses.

**Isolated Stellar Mass Black Holes**

Stars with initial masses of ~8 to 20 M☉ end their lives as neutron stars (NSs), and the more massive as black holes (BHs) (Heger et al. 2003). For a reasonable range of ages, IMFs, and stellar-wind characteristics, as much as 10% of the total mass of an old stellar population should be in the form of NSs and BHs (Oslowski et al. 2008). Up to 99% of massive stellar remnants are expected to be single, either primordially or due to disruption of binaries by supernova (SN) explosions (Agol & Kamionkowski 2002). Such isolated stellar-mass black holes (ISMBHs) are extremely difficult to detect directly, as they emit no light and accrete from the ISM at very low rates.

ISMBHs can be best detected through microlensing. Roman is expected to detect several thousand microlensing events towards the Galctic bulge. If stellar remnants constitute a few percent of the total mass, some of these observed microlensing events will be due to ISMBHs which will reveal themseves as long-duration microlensing events. However, microlensing light curves are degenerate with respect to the mass and transverse velocity of the lens (and, more weakly, the distance to the lens). Thus a long-duration event with no light contribution from the lens could arise from a high-mass, non-luminous NS or BH lens, but it could also be due to an unusually slow-moving low-mass stellar lens.

A route to resolving these degeneracies arises from the fact that microlensing, in addition to amplifying the brightness of the source, produces a small shift in its position (Dominik and Sahu, 2000, Figure 1). Thus if high-precision astrometry is added to the photometry, the deflection of the source image can be measured, and thus the mass of the lens determined unambiguously. This was recently demonstrated by Sahu et al. (2022) who reported the first unambiguous detection of an ISMBH through this technique. Sahu et al. (2022) selected long-duration events detected by ground-based monitoring programs, and used HST to measure the small deflections (~milliarcsec) caused by the lens. *Roman* will not only detect microlensing events caused by ISMBHs, but will also simultaneously measure the tiny deflections and thus measure their masses.

**Current Observational Strategy:**

*Roman*'s bulge microlensing program has been planned to detect mostly short-duration microlensing events due to exoplanets beyond the snow line of main-sequence stars and free-floating exoplanets. Nevertheless, the duration of its mission is long enough to detect and characterize long-duration microlensing events, and its astrometric accuracy is high enough to discern the astrometric trajectories of source stars.

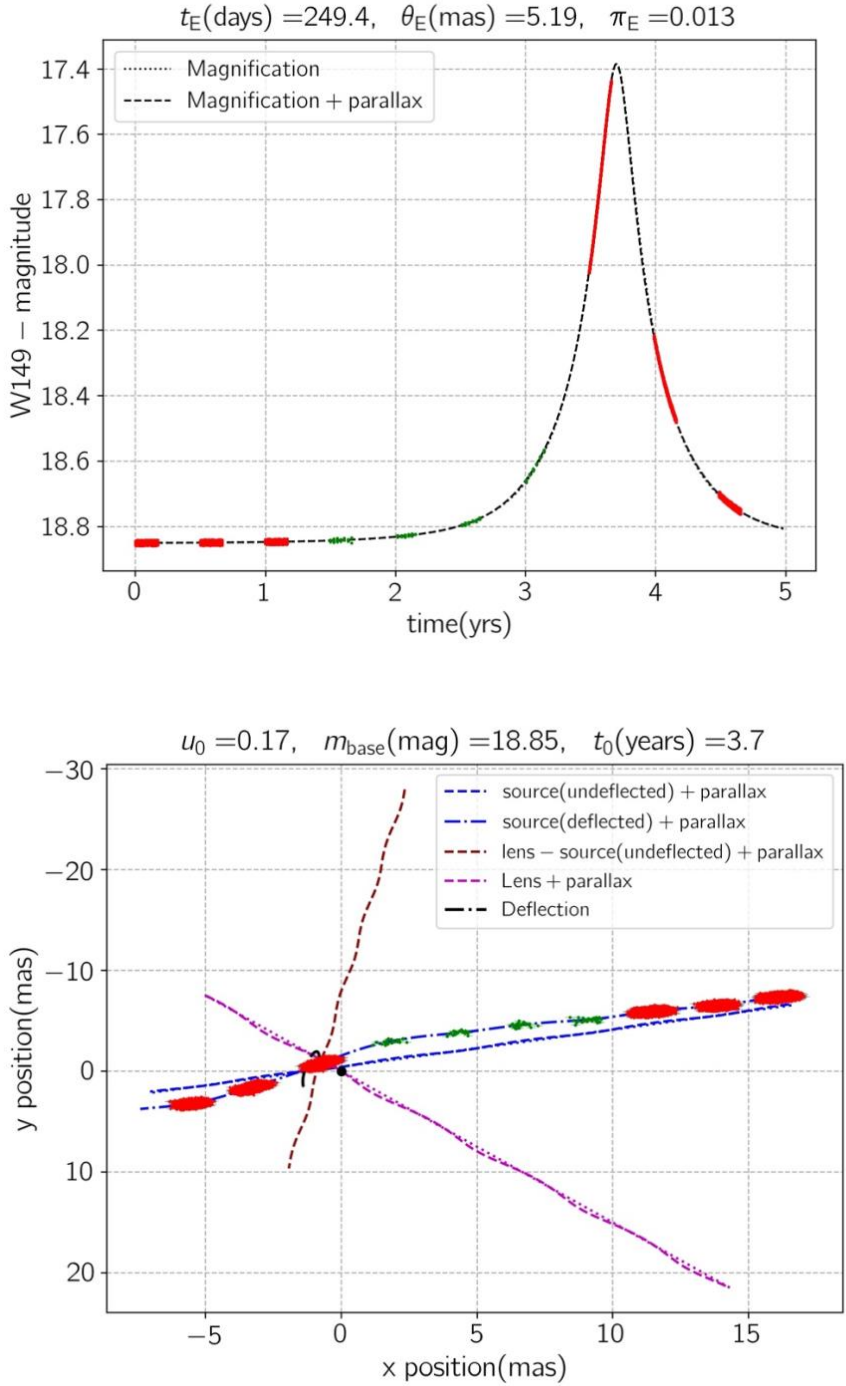

Fig 1. An example of a simulated microlensing event, as observed by *Roman*. The top panel shows the magnification curve with (dashed curves) and without (dotted curves) the parallax effect. The bottom panel shows the corresponding astrometric motion of the source star (blue curve), lens object (magenta curve), and their relative motions (dark red curve) projected on the sky plane. The long time gap in the currently scheduled Roman observations leads to degenerate solutions. The green dots represent the small amount of additional observations proposed here, which make the detection robust, and provide a firm mass estimate of the ISMBH.

Unfortunately, the currently planned long time gap between *Roman*'s first three observing seasons and the other three seasons would limit its efficiency and robustness for discerning and characterizing ISMBHs. This is demonstrated in Figure 1, which shows a simulated long-duration microlensing event. The top panel shows the magnification curve with (dashed curves) and without (dotted curves) the parallax effect. The bottom panel shows the corresponding astrometric motion of the source star (blue curve), lens object (magenta curve), and their relative motions (dark red curve) projected on the sky plane. The synthetic data are taken with the *Roman* telescope. The long time gap in the currently scheduled Roman observations leads to degenerate solutions.

**Recommended Strategy:**

As shown by Sajadian and Sahu (2023), the situation can be greatly improved by adding one hour of observations (4 data points) every ~10 days when the Galactic bulge is detectable in our simulations. These additional observations amount to a total of about one to two days of observations with Roman. The additional observations are shown as green points in Fig.1. This small amount of extra observations improve the situation and make the solution unique. We thus recommend *a small amount of additional observations –one hour of observations every ~10 days when the Bulge is observable during the large time gap– which is equivalent to a total of about one to two additional days of observations with Roman.* The observations should use the same filters and integration times as the survey program. This small amount of additional observations will greatly improve the efficiency and robustness of detection of ISMBHs, and provide firm estimates of their masses.

**References**

Agol, E., & Kamionkowski, M. 2002, MNRAS, 334, 553
Dominik, M., & Sahu, K. C. 2000, ApJ, 534, 213
Fryer, C., & Kalogera, V. 2001, ApJ, 554, 548
Heger, A., et al. 2003, ApJ, 591, 288
Oslowski, S., et al. 2008, A&A, 478, 429
Sajadian, S, & Sahu, K.C., 2023, Astron. J. 165, 96
Sahu, K. et al. 2022, ApJ, 993, 83